\documentclass[twocolumn,amssymb,amsmath,aps,prl,groupedaddress,nobibnotes,showpacs]{revtex4}
\usepackage{graphicx}

\begin{document}
\title{Can quantum imaging be classically simulated?}
\author{Milena D'Angelo, and Yanhua Shih}
\affiliation{Department of Physics, University of Maryland, Baltimore
County, Baltimore, Maryland 21250}

\date{24 February, 2003}

\begin{abstract}
Quantum imaging has been demonstrated since 1995 by using
entangled photon pairs.  The physics community named these
experiments ``ghost image", ``quantum crypto-FAX", ``ghost
interference", etc. Recently, Bennink et al simulated the ``ghost"
imaging experiment by two co-rotating {\bf k}-vector correlated
lasers. Did the classical simulation simulate the quantum aspect
of the ``ghost" image?  We wish to provide an answer. In fact, the
simulation is very similar to a historical model of local realism.
The goal of this article is to clarify the different physics
behind the two types of experiments and address the fundamental
issues of quantum theory that EPR was concerned with since 1935.
\end{abstract}


\maketitle

A series of nonlocal quantum imaging experiments have been
realized, since 1995, using entangled photon pairs generated in
spontaneous parametric down conversion  (SPDC)
\cite{Ghost_d},\cite{Ghost_i}, \cite{Popper_exp}. The physics
community named them ``ghost image", ``quantum crypto-FAX",
``ghost interference", etc. Those experiments have demonstrated
the peculiar behavior of entangled states. The insertion of an
optical element, mask or single-double-multi-slit in the arm in
which one of the entangled photon propagates, allows observing
either an image (in the position defined by two-photon Gaussian
thin lens equation \cite{Ghost_i}) or the Fourier transform of the
object (in the far field zone \cite{Ghost_d}) when joint
detections of the pair are registered. However, neither of the
single detections in either arm is able to reconstruct the image
or to give the Fourier transform of the object. This effect has
later been used in a two-photon Young's interference-diffraction
experiment to demonstrate the working principle of quantum
lithography \cite{lithography}, \cite{Dowling}. By measuring the
interference-diffraction pattern on the Fourier transform plane,
we showed that the spatial resolution of a two-photon image could
be improved by a factor of 2, beyond the diffraction limit. In
fact, a two-photon entangled state at wavelength $\lambda$
produces a Fourier transform pattern equivalent to the one given
by classical light (single photon) at wavelength $\lambda/2$. This
effect was well explained by considering the entangled nature of
the photon pair, i.e., the coherent superposition of the
two-photon amplitudes.

Recently, Bennink {\em et al.} simulated the ``ghost" imaging
experiment using two classically {\bf k}-vector correlated pulses
emitted by co-rotating lasers \cite{Bennink}. An object is
inserted in the arm in which one of the pulses propagates while
the other pulse propagates freely in a separate arm; by recording
coincidences between pulses of each pair, shot by shot, this
experiment has also reconstructed an ``image" of the object in the
joint detections.

Has the recent classical experiment simulated the quantum aspects
of the ``ghost" image? We wish to provide an answer in this
article by analyzing and comparing the ``ghost" image experiment
with its recent classical simulation. In fact, the simulation is
very close to a historical model of local realism for an entangled
two-particle system.  Our goal is to clarify the very different
physics behind the two types of experiments and address the
fundamental issues of quantum theory that Einstein-Podolsky-Rosen
was concerned with since 1935 \cite{EPR}. In addition, important
practical advantages of quantum imaging will be emphasized.

\begin{figure}[t]
\includegraphics[width=3.0in]{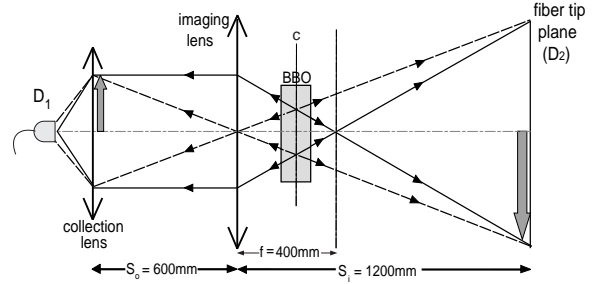}
\caption{\label{unfold_qu}Unfolded version of the two-photon
``ghost" imaging experiment. The SPDC source (BBO crystal) emits
bi-photon wave packets: the momentum-momentum correlation of the
two-photon system is defined with certainty but the momentum of
the signal and the idler separately is completely undefined. The
image of the object is the result of a click-click joint detection
of a pair. The straight lines correspond, schematically, to all
possible two-photon amplitudes. Note that even if they have
different propagation directions, they are indistinguishable for a
joint detection event. This allows the possibility for the imaging
lens to create an image of the object at position $S_i$ such that:
$1/S_o + 1/S_i = 1/f$. The spatial resolution is limited by
two-photon diffraction, as resulting from the superposition of the
two-photon amplitudes.}
\end{figure}

\begin{figure}[t]
\includegraphics[width=3.0in]{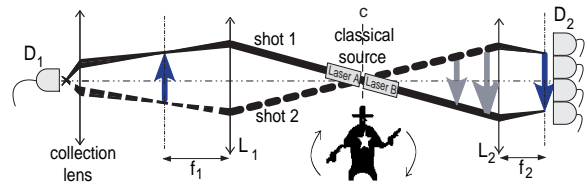}
\caption{\label{unfold_cl}Unfolded version of the classical
simulation using two laser beams. A well-defined momentum
characterizes each laser beam. The apparatus works as a projector:
a projection of the object is formed, shot by shot, anywhere on
the right side of the source. There is no need of an imaging lens
and no lens-image equation to satisfy. The use of lenses is for
reducing the spot size of the laser beams, which determines the
spatial resolution of the projection. As a matter of fact, two
correlated guns could give the same result.}
\end{figure}

To compare two-photon ``ghost" imaging with its classical
simulation, let us consider the ``unfolded" version of both the
``ghost" image experiment, as realized in Ref. \cite{Ghost_i}, and
its classical simulation, as realized in Ref. \cite{Bennink}. The
schematic experimental diagrams are shown in Fig.~\ref{unfold_qu}
and Fig.~\ref{unfold_cl}, respectively.

In the quantum two-photon imaging experiment a pair of
signal-idler photon is generated in the process of SPDC inside a
nonlinear crystal (BBO). The state of the signal-idler pair can be
calculated by the first order perturbation theory and has the form
\cite{Klyshko}:

\begin{equation}\label{state_SPDC}
\begin{array}{c}
\left| \Psi \right\rangle =\int{d{\bf k}_i,d{\bf k}_s} \delta({\bf
k} _s+{\bf k} _i-{\bf k}_p) \delta(\omega _s+\omega _i-\omega
_p)\\a_{s}^{\dagger }({\bf k}_{s})\ a_{i}^{\dagger }({\bf
k}_{i})\mid 0\rangle
\end{array}
\end{equation}

where $\omega_j$, $\bf k_j$ ($j = s, i, p$) are the frequency and
wavevector of the signal, idler and pump, respectively, and
$a_{s}^{\dagger }$ and $a_{i}^{\dagger }$ are the creation
operators for the signal and idler photons, respectively.  The
energy and momentum for neither signal nor idler photon is
defined. However, the energy-energy and momentum-momentum
correlations of the pair are defined with certainty, as expressed
by the delta functions, which are technically called phase
matching conditions:
\begin{equation}  \label{phasematch}
\omega _s+\omega _i=\omega _p,\quad {\bf k}_s+{\bf k}_i={\bf k}_p,
\end{equation}
In the degenerate case, $\omega _s=\omega _i$, the transverse wave
vector phase matching requires the signal and idler belonging to
one pair to be emitted at equal, yet opposite, angles relative to
the pump. In other words, the propagation direction for either
photon may have a great uncertainty, but the correlation of the
emission angle is determined with certainty. This then allows for
a simple pictorial viewing of the experiment in terms of ``usual''
geometric optics, treating the SPDC crystal as a ``mirror". This
concept has been simplified in Fig.~\ref{unfold_qu} by drawing
straight lines representing the probability amplitudes, or the
quantum ``pathways", associated with a signal-idler pair. Note
that: 1) each straight line represents a two-photon amplitude,
defining a {\em possible} special momentum-momentum correlation
with a defined propagation direction of the signal-idler pair; 2)
all the two-photon amplitudes belong to {\em one pair} of
signal-idler photon, they exist {\em simultaneously and are
indistinguishable}. The insertion of an optical lens (``imaging
lens" in Fig.~\ref{unfold_qu}) in the signal side allows the
two-photon amplitudes to make an image of the object on the idler
side. The image appears in coincidence measurement between $D_1$
and $D_2$, while single counts on $D_1$ and $D_2$ are both
constant. The location of the object plane and the image plane is
defined by the two-photon Gaussian thin lens equation: $1/f =
1/S_i + 1/S_o$, with $f$, $S_i$ and $S_o$ as defined in
Fig.~\ref{unfold_qu}. Indeed, from the geometric point of view,
the image and object are related by a precise point-to-point
correspondence, which represents the position-position correlation
of the pair.  In principle, the entire image of the entire object
is formed by the two-photon probability amplitudes of {\em one
pair}. All the two-photon amplitudes that end on the object plane
and the image plane are indistinguishable, which allows the image
to be coherent. The quantum image is then very special: (1) it is
useful for some cryptography-type applications; (2) it may be
sub-diffraction limited, since its spatial resolution is
determined by the superposition of the two-photon amplitudes.

On the other hand, the classical simulation represented in
Fig.~\ref{unfold_cl}, employs two co-rotating {\bf k}-vector
correlated laser beams (pulses), each of them pointing in a {\em
well defined direction}. The pulses ``know" where to go in the
course of their propagation. Each pair of pulses can be focalized
at a defined point on the object and the ``image" plane with the
help of lenses $L_1$ and $L_2$, respectively.  Consequently, a
point on the object plane is {\em projected} onto the ``image"
plane at the CCD camera by recording ``coincidences" shot by shot.
This shot by shot, point-to-point projection works like a ``two
way" ``Chinese shadow".  It should be emphasized that one pair of
laser pulses can project only one point and each projection event
is well distinguishable from the others. Consequently, the result
of this experiment is a {\em collective projection}. The spatial
resolution of the projection is determined by the spot size of the
laser beam on the object and ``image" plane.  The use of lenses
$L_1$ and $L_2$ is for reducing the spot size of the laser beam.
In principle, $L_1$ and $L_2$ are not necessary for the projection
to be realized.  There is no general lens-image equation to
satisfy: the ``image" plane is independent of the object plane and
is fixed only by the position of $L_2$, which can be inserted
everywhere on the right side. As a matter of fact, two correlated
guns could give the same result.

It is clear at this point that the two-photon ``ghost" image
experiment works as a coherent imaging system
(Fig.~\ref{unfold_qu}) while the classical simulation works as a
shot by shot, point-to-point projector (Fig.~\ref{unfold_cl}).

In the above picture of two-photon entanglement, the entangled
two-photon state of SPDC and the two-photon probability amplitudes
play the fundamental role. Indeed, to simulate the quantum aspect,
one needs to simulate the coherent superposition of the two-photon
amplitudes. In principle, the simulation should produce the entire
object-image point-to-point correlation with one pair of laser
pulses instead of shot-by-shot.

We may have to answer the following questions: Are we sure the
above picture of quantum entanglement is physically true? And
since in both experiment the final result is obtained by counting
coincidences between either two-photon or two pulses, are we sure
the real physical process in the measurement of the``ghost" image
is not the same as in the classical simulation? In fact, the
picture of quantum entanglement has never been accepted by local
realism. Since 1935, Einstein-Podolsky-Rosen \cite{EPR} were
seriously concerned with the physics of entangled two-particle
system and questioned it. The ``ghost" image appears as a
``nonlocal" effect, which EPR \cite{EPR} consider as an absurd
``action-at-a-distance"; they reject this idea and conclude that
when the pair is created it has to be predetermined for it ``where
to go", just like in the classical simulation.

We will try to answer the above questions in the following way:
(1) confirm that, in the two-photon image-type experiment, the
two-photon probability amplitudes play the fundamental role; (2)
show that, the classical simulation can never simulate the
two-photon probability amplitudes and their coherent
superposition; (3) demonstrate an unique advantage of two-photon
coherent imaging, which, in principle, can never be realized by
classical simulations.

\begin{figure}[t]
\includegraphics[width=3.0in]{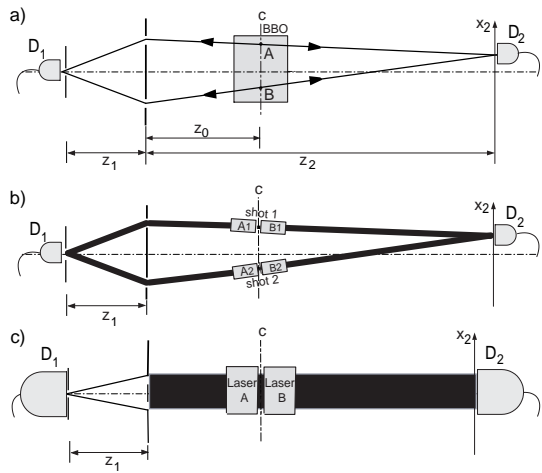}
\caption{\label{Interference}a) Unfolded version of the two-photon
``ghost" interference experiment. The two-photon Fourier transform
of the double slit is the result of a click-click joint detection
of an entangled photon pair. b) A pair of co-rotating laser beams
is used to simulate classically the two-photon amplitudes, shot by
shot. c) Another approach for the classical simulation. A single
pair of lasers is used to cover simultaneously both the upper and
the lower slit.}
\end{figure}

Is it true that the two-photon probability amplitudes play the
fundamental role?  One may find the answer studying the spatial
resolution of the ``ghost" image. The spatial resolution of an
image, assuming the use of perfect lenses, is basically determined
by diffraction or equivalently by the uncertainty relations
\cite{Feynman}.  If one could show that the diffraction effect in
a two-photon imaging type experiment is due to the superposition
of the two-photon amplitudes, it would be possible to conclude
that the two-photon amplitudes are also responsible for the
formation of the image itself. In order to study the spatial
resolution of an image it may be easier to move our attention from
the image plane to the Fourier transform plane of the object. The
sister experiment of the ``ghost" image, which received the name
of ``ghost interference" \cite{Ghost_d}, may serve this purpose.
The experimental setup was very similar to the ``ghost" image
experiment and used the same two-photon source of entangled state
(SPDC). Fig.~\ref{Interference}a is an unfolded version of the
schematic experimental setup.  A Young's double-slit serves as a
``complicated pattern" - the object.  The measurement was done by
counting coincidences in the far-field-zone to study the
two-photon Fourier transform of the double-slit.

The experimental result is very surprising from the viewpoint of
classical physics. There is no interference pattern behind the
double-slit. However, the joint detections reproduce an
interference-diffraction pattern when detector $D_2$ is scanned
across the idler beam while $D_1$ is fixed. The lack of first
order interference is due to the poor spatial coherence of the
signal beam, i.e., the diverging angle of the beam for a given
wavelength $\lambda$ is greater then $\lambda /d$ where $d$ is the
spacing between the double-slit.  The two-photon
interference-diffraction pattern is predicted by quantum
entanglement theory and the prediction agrees with the
experimental result. The interference-diffraction pattern which
appears measuring coincidences is:
\begin{equation}  \label{RcD2}
R_c(x_2)\propto sinc^2(x_2\pi a/\lambda z_2)\cos^2(x_2\pi
d/\lambda z_2)
\end{equation}
where $a$ is the slit width, $\lambda$ is the central wavelength
of signal and idler, $x_2$ is the coordinate of detector $D_2$.
This is a standard Young's double slit interference-diffraction
pattern, except for the fact that $z_{2}$ is the distance from the
double-slit, back to the SPDC crystal, and then forward to the
scanning detector $D_2$ (in analogy to the definition of $S_i$ in
the ``ghost" image). It is straightforward to calculate the
pattern of Eq.~\ref{RcD2} from the entangled two-photon state of
SPDC. To simplify the discussion, let us consider the interference
first. The probability of joint detections is proportional to the
norm squared of $<0|E_{1}^{+} E_{2}^{+}|\Psi>$, where $E_{1}^{+}$
and $E_{2}^{+}$ are the field operators at detectors $D_1$ and
$D_2$, respectively.  As schematically represented by the straight
lines in Fig.~\ref{Interference}a, only the two-photon amplitudes
that pass through the double-slit can give rise to coincidences.
So we may write $|\Psi>$ as a simplified version of the general
SPDC state (eq.~\ref{state_SPDC}), considering only the four mode
state vector:
\begin{equation}\label{state_SPDC_slit}
\left| \Psi \right\rangle =\varepsilon[a_{s}^{\dagger }
a_{i}^{\dagger } +b_{s}^{\dagger } b_{i}^{\dagger }]\mid 0\rangle,
\end{equation}
where $\varepsilon$ is a constant, $a_{j}^{\dagger }$
($b_{j}^{\dagger }$) is the photon creation operator for the upper
(lower) mode in Fig.~\ref{Interference}a ($j = s, i$).
Substituting the field operators and the state vector into
$<0|E_{1}^{+} E_{2}^{+}|\Psi>$, the probability of joint
detections is then given by the cosine function in Eq.~\ref{RcD2}.
The diffraction pattern can be calculated by integrating the many
two-photon amplitudes over the slit width ($-a/2 < x < a/2$). The
interference-diffraction is the result of the coherent
superposition of the two-photon amplitudes of a signal-idler pair.
The straight lines in Fig.~\ref{Interference}a are responsible for
the two-photon Fourier transform of the double-slit aperture
function, and consequently, for the formation of the image in
Fig.~\ref{unfold_qu}.

Can the two-photon ``ghost" interference-diffraction be simulated
classically?  Let us try two different approaches.

First, assume one can find a way to simulate the two amplitudes of
Fig.~\ref{Interference}a, for example by means of two pairs of
laser pulses, as depicted in Fig.~\ref{Interference}b. Pulse A
passes the upper slit and the lower slit in two different shots,
similarly to the ``shot-by-shot" operation in ref.~\cite{Bennink}.
It is clear that there will be no interference, not even in
principle. The ``upper shot" and the ``lower shot" are well
distinguishable. What about diffraction? Pulse A itself will
experience diffraction when passing the upper and the lower slit
separately. In the single counts on $D_1$ there will be two
diffraction patterns shadowing each other. However, these
diffraction patterns have nothing to do with pulse B.  The Fourier
transform of the double-slit function is done only by laser pulse
A.  This is fundamentally different from the two-photon ``ghost"
interference-diffraction.

The second approach is illustrated in Fig.~\ref{Interference}c.
Laser beam A covers both the upper and the lower slit.  In this
case, one shot of laser pulse A will make a first order standard
Young's interference-diffraction pattern.  Again, it has nothing
to do with laser pulse B.

Unless the two-photon amplitudes and the two-photon entangled
state are simulated, it is impossible to obtain the two-photon
Fourier transform of the double-silt.

An unique advantage of using two-photon entangled states for
imaging is the improvement of the image spatial resolution, even
beyond the diffraction limit.  This is a hot topic of quantum
lithography.  Recently, we have realized a Young's double slit
experiment to demonstrate the working principle of quantum
lithography \cite{lithography}.

\begin{figure}[t]
\includegraphics[width=3.0in]{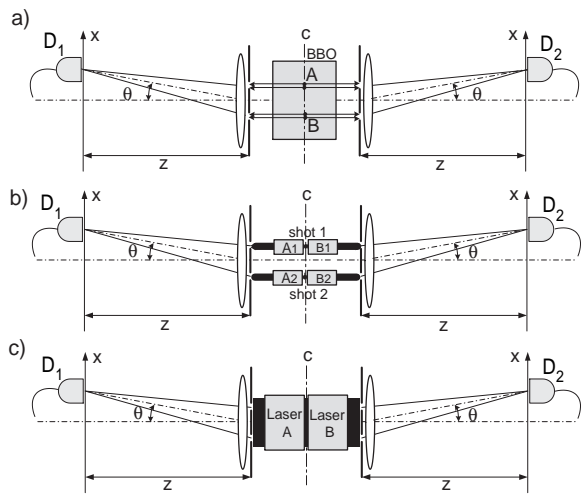}
\caption{\label{litho}a) Unfolded version of the quantum
lithography experiment. b) Classical simulation, shot-by-shot,
using a pair of lasers. c) Another approach of classical
simulation: both laser beams cover the double-slit
simultaneously.}
\end{figure}

The philosophy of our experiment is to show that the N-photon
Fourier transform of an object at wavelength $\lambda$ is
equivalent to the Fourier transform obtained using classical light
at wavelength $\lambda/N$. This would prove that the spatial
resolution of the reduced-size image obtained by a second set of
lenses would be N times better. Fig.~\ref{litho}a is the unfolded
version of the simplified experimental setup. In our experiment, a
Young's double-slit played the role of the ``complicated" pattern.
By using two-photon entangled states emitted from SPDC under
certain experimental conditions, we found that the two-photon
double-slit interference-diffraction pattern, in the far field
zone, has modulation period of the spatial interference smaller,
and width of the diffraction pattern narrower, both by a factor of
two, than those of the classical case. This means that the Fourier
transform for the entangled two-photon light at wavelength
$\lambda$ is equivalent to the one obtained using classical light
at $\lambda/2$, instead of $\lambda$. The physics and the
mathematics are similar to those of the ``ghost" interference
experiment. Indeed, under certain experimental conditions, we
managed to have the two-photon amplitudes always passing through
one slit (upper or lower), as depicted in Fig.~\ref{litho}a. The
two-photon interference-diffraction is calculated as superposition
(integration) of those two-photon amplitudes over the double slit,
which is:
\begin{equation}  \label{P12}
P(x)\propto{\rm sinc}^{2}[(2\pi a/\lambda)x/z]  {\rm
cos}^{2}[(2\pi d/\lambda)x/z],
\end{equation}
where $z$ is the distance between the slit and the ``two-photon"
detector, and $x$ ($x =x_1 = x_2$) is the coordinate of the
``two-photon" detector measured from the center of the pattern.

What happens if one replaces the two-photon source (SPDC) by a
pair of {\bf k}-vector correlated laser beams, as Bennink {\em et
al.} did in Ref. \cite{Bennink}? The situation is shown in
Fig.~\ref{litho}b,c. It is easy to see that each laser beam,
independently, may produce a ``classical" Fourier transform of the
slit. As a consequence, on the image plane, the final image would
be a ``product" of the two. And since each one of them,
separately, must be subject to the classical diffraction limit,
the final image (obtained by joint detections) must also be
subject to it. The classical simulations cannot improve the
spatial resolution of an image beyond the diffraction limit. It
would really be a violation of the uncertainty principle if it
did.

Conclusion: quantum imaging experiments have explored the very
special physics of multi-particle entanglement. The multi-particle
probability amplitudes play the fundamental role in quantum
imaging. In general, they do not have a classical analogous and,
consequently, cannot be simulated classically. The coherent
superposition of the multi-particle amplitudes result in effects
which are unacceptable by classical physics. An interesting
alternative demonstration of the quantum character of entangled
imaging has been recently proposed in \cite{Gatti}. Feynman used
to consider the superposition principle as the only mystery of
quantum mechanics \cite{Feynman}. Indeed, in an quantum entangled
system, superposition takes place in a special form: it is the
superposition of multi-particle amplitudes.  One of the
consequences of quantum superposition is the effect of
sub-diffraction (or super-resolution). As Feynman pointed out in
his Lectures \cite{Feynman} the effect of diffraction reflects the
same physics contained in the uncertainty principle
\cite{Popper_exp}. It follows that the uncertainty relations, as
EPR \cite{EPR} and Popper \cite{Popper} argued in 1935, may serve
as another standard, besides Bell inequality \cite{Bell}, for
distinguishing quantum from classical physics.

The authors thank M.H. Rubin for helpful discussions.  This work
was supported, in part, by ONR and NSF.

\end{document}